\documentclass{PoS}
\usepackage{subfig}
\usepackage{array}
\usepackage{dcolumn}
\newcolumntype{d}[1]{D{.}{.}{#1} }
\newcolumntype{?}{!{\vrule width 3pt}}

\title{Probing Convolutional Neural Networks for Event Reconstruction in $\gamma$-Ray Astronomy with Cherenkov Telescopes}

\ShortTitle{Probing CNNs for Event Reconstruction in $\gamma$-Ray Astronomy with IACTs}

\author{\speaker{Tim Lukas Holch} $^a$, Idan Shilon $^b$, Matthias B\"uchele $^b$, Tobias Fischer $^b$, Stefan Funk $^b$, Nils Groeger $^a$, David Jankowsky $^b$, Thomas Lohse $^a$, Ullrich Schwanke $^a$, Philipp Wagner $^a$\\
	\llap{$^a$} Humboldt University of Berlin, Newtonstr. 15, 12489 Berlin, Germany \\
	\llap{$^b$} Friedrich-Alexander University Erlangen-N\"urnberg, Erwin-Rommel-Str. 1, 91058 Erlangen, Germany\\	
    E-mail: \email{Tim.Holch@physik.hu-berlin.de}}

\abstract{A dramatic progress in the field of computer vision has been made in recent years by applying deep learning techniques. State-of-the-art performance in image recognition is thereby reached with Convolutional Neural Networks (CNNs). CNNs are a powerful class of artificial neural networks, characterized by requiring fewer connections and free parameters than traditional neural networks and exploiting spatial symmetries in the input data. Moreover, CNNs have the ability to automatically extract general characteristic features from data sets and create abstract data representations which can perform very robust predictions. This suggests that experiments using Cherenkov telescopes could harness these powerful machine learning algorithms to improve the analysis of particle-induced air-showers, where the properties of primary shower particles are reconstructed from shower images recorded by the telescopes. In this work, we present initial results of a CNN-based analysis for background rejection and shower reconstruction, utilizing simulation data from the H.E.S.S. experiment. We concentrate on supervised training methods and outline the influence of image sampling on the performance of the CNN-model predictions.}

\FullConference{35th International Cosmic Ray Conference --- ICRC2017\\
		10--20 July, 2017\\
		Bexco, Busan, Korea}

\begin{document}

\section{Introduction}

Ground-based $\gamma$-ray astronomy opens a window to the TeV energy range and provides a picture of some of the most exciting processes in the Universe. Many observations in this so-called very high energy (VHE; E > 100 GeV) regime are performed with Imaging Atmospheric Cherenkov Telescopes (IACTs). These telescopes are able to detect VHE cosmic $\gamma$-rays and cosmic rays (CRs) by imaging the Cherenkov light emitted by so-called extensive air showers (EASs), the result of the primary particle's interaction with nuclei in the Earth's atmosphere. Images of EASs can provide information to determine the primary particle's type and energy, and for $\gamma$-rays also the direction.

The analysis of EASs improves significantly when observing the showers from several angles. A state-of-the-art array of IACTs that utilizes this stereoscopic approach is the High Energy Stereoscopic System (H.E.S.S.)~\cite{hess_1}. H.E.S.S. is located in the Khomas highland of Namibia and comprises five IACTs of two different sizes, four telescopes (CT1-4) with a reflector area of $\sim 100\,$m$^2$ and one larger telescope (CT5) with $\sim 600\,$m$^2$.

To successfully analyse data taken with IACTs, one must separate the $\gamma$-ray induced signal from the dominating background of hadron-induced showers, and accurately reconstruct the $\gamma$-rays' direction and energy. Current widely used reconstruction techniques rely on the so-called Hillas parameters~\cite{hillas}, the moments of the measured intensity distributions in individual camera images. More advanced approaches use likelihood fits to idealized semi-analytical shower models~\cite{model} or templates from Monte-Carlo (MC) simulations~\cite{impact} to improve the reconstruction accuracy. Furthermore, by introducing traditional machine learning concepts, e.g. boosted decision trees, to the background suppression stage of the analysis, the performance of these analysis methods was significantly improved~\cite{hess_bdt}.

In its core, the analysis of IACT data relies on deciphering the EAS images. Hence, utilising modern, machine learning based, image-analysis techniques seems like the natural next step in the evolution of IACT analysis tools. A particular type of such a technique, the Convolutional Neural Networks (CNNs), was specifically designed to address image recognition tasks and has been showing impressive performance in supervised recognition and inference problems~\cite{imagenet_cnns}. In this work, we apply CNNs to H.E.S.S. MC simulation data and study its application to background suppression and event reconstruction.


\section{Convolutional Neural Networks} 
\label{sec_cnns}

A CNN is a specialized kind of artificial neural network for processing data of grid-like structure~\cite{dl_cnn}. CNNs take a complete grid of image pixels as inputs (where in our case the grid is two-dimensional). Usually, their architecture includes numerous convolutional layers (CLs), followed by a number of fully connected layers (FCs). A CL typically comprises three stages. In the first stage of a CL, a convolution is performed between a learnable kernel and the input image, yielding linear activation outputs. The kernel size is smaller than the input size (e.g. $5 \times 5$) and allows to detect meaningful features (e.g. edges of a shape) which occupy only a small part of the image. This {\it sparse connectivity} reduces the number of free parameters and memory requirements of the model compared to traditional neural networks. Moreover, each kernel parameter is used (or {\it shared}) at every position of the image (disregarding boundary effects). Such {\it parameter sharing} further reduces memory usage. Parameter sharing also implies that CNNs are {\it equivariant} to translations, meaning that a shift in the input leads to the same shift in the output.

The convolution stage of a CL is followed by an activation stage, where each linear activation is fed into a nonlinear activation function such as the rectified linear unit~\cite{relu}. The third stage is called the pooling stage. A pooling function~\cite{pool} replaces the output of the layer at a certain location with a summary statistic of the nearby outputs, such as the maximum output in the neighbourhood of that location. The pooling operation makes the output of the convolution become approximately invariant to small translations in the input. Together, the above mentioned CL features make CNNs excel at efficient computer object recognition.


Usually, one normalizes the output of each CL to improve the network's optimization. Typical normalizations include local response normalization~\cite{lrn} and batch normalization~\cite{bn}. To address overfitting, the FC layers can include regularization strategies, such as dropout~\cite{dropout} and weight decay. The last layer of the network is a simple linear layer. To classify inputs, one can feed the linear inputs to a softmax function to yield a probabilistic measure and predict the class of the input image. To update the weights, a cross-entropy loss is often used. For regression, the linear outputs are fed into a mean squared error. To implement our models, incorporating the CNN features above, we have been utilizing the Caffe~\cite{caffe} and TensorFlow~\cite{tf} deep learning (DL) frameworks.



\section{Image Pre-Processing}
\label{sec_data}

\begin{figure}
\centering
\includegraphics[width=0.22\textwidth]{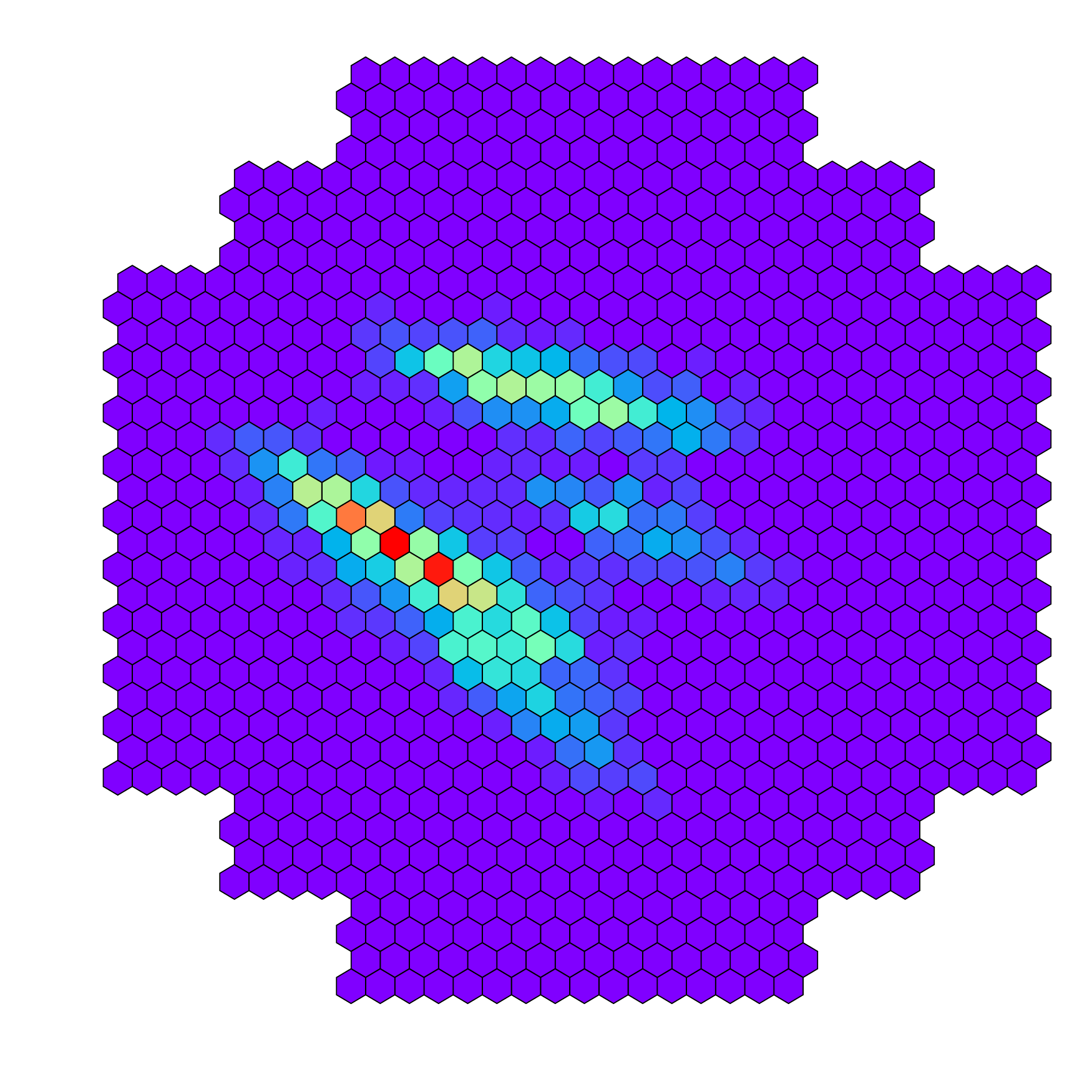}
\includegraphics[width=0.22\textwidth]{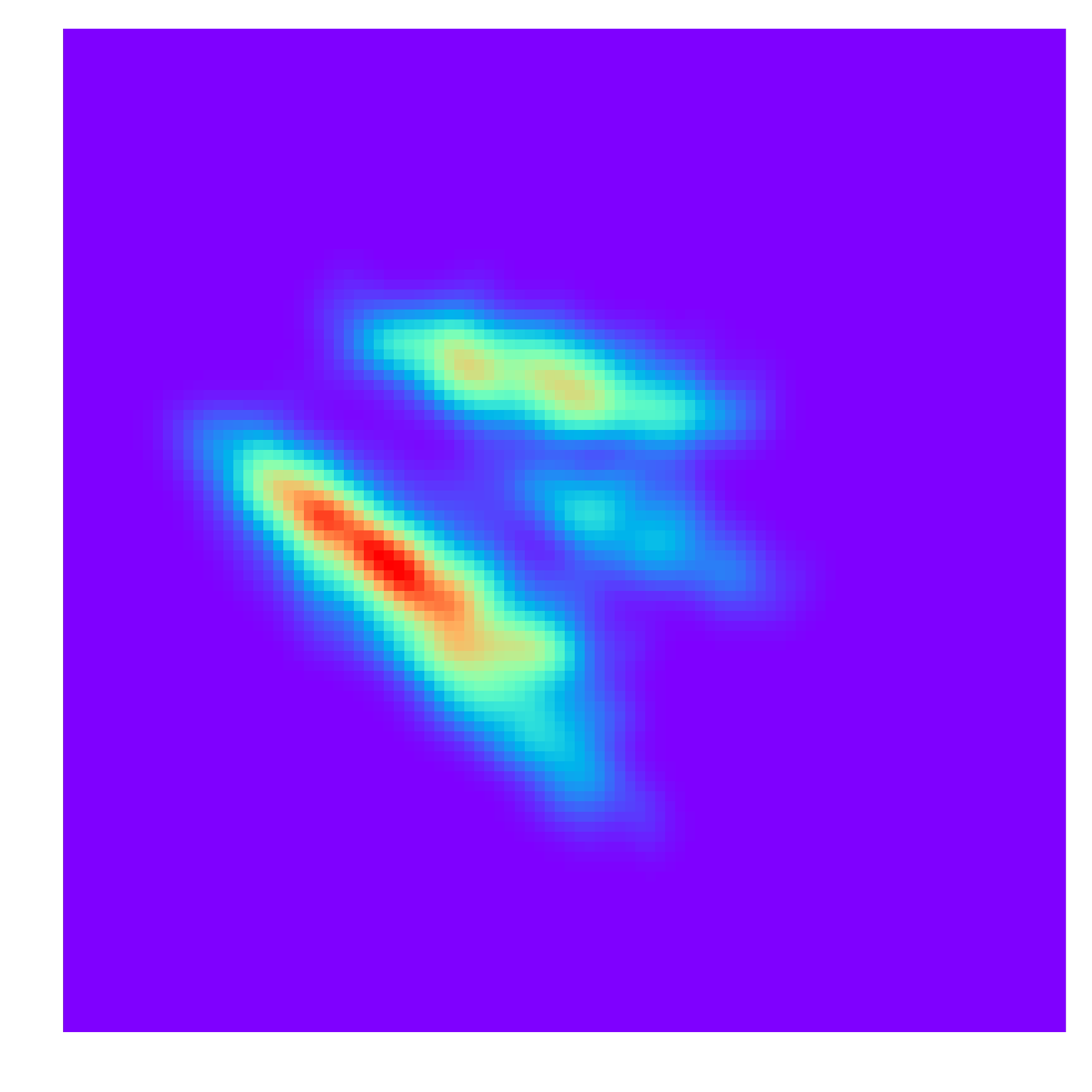}
\includegraphics[width=0.22\textwidth]{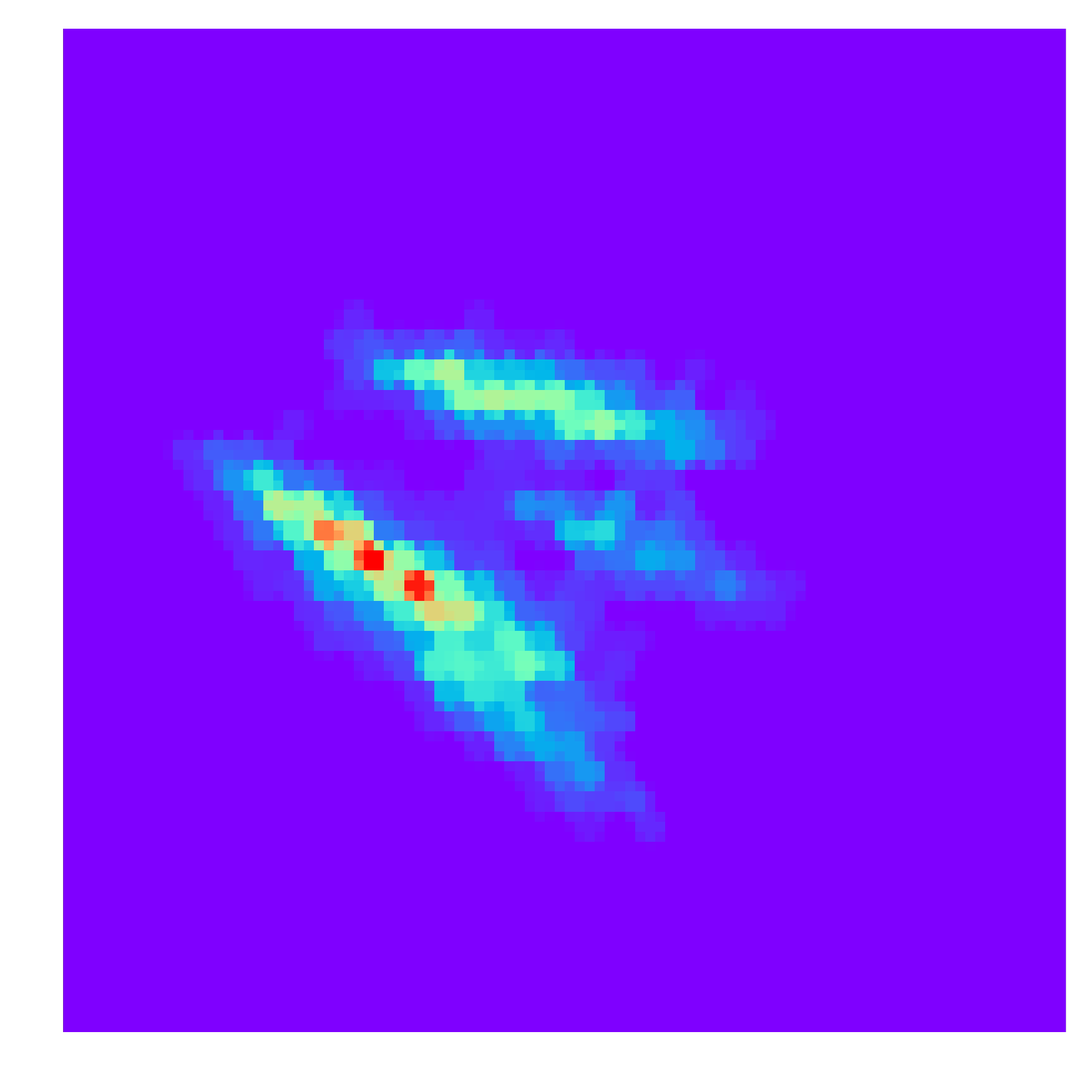}
\includegraphics[width=0.22\textwidth]{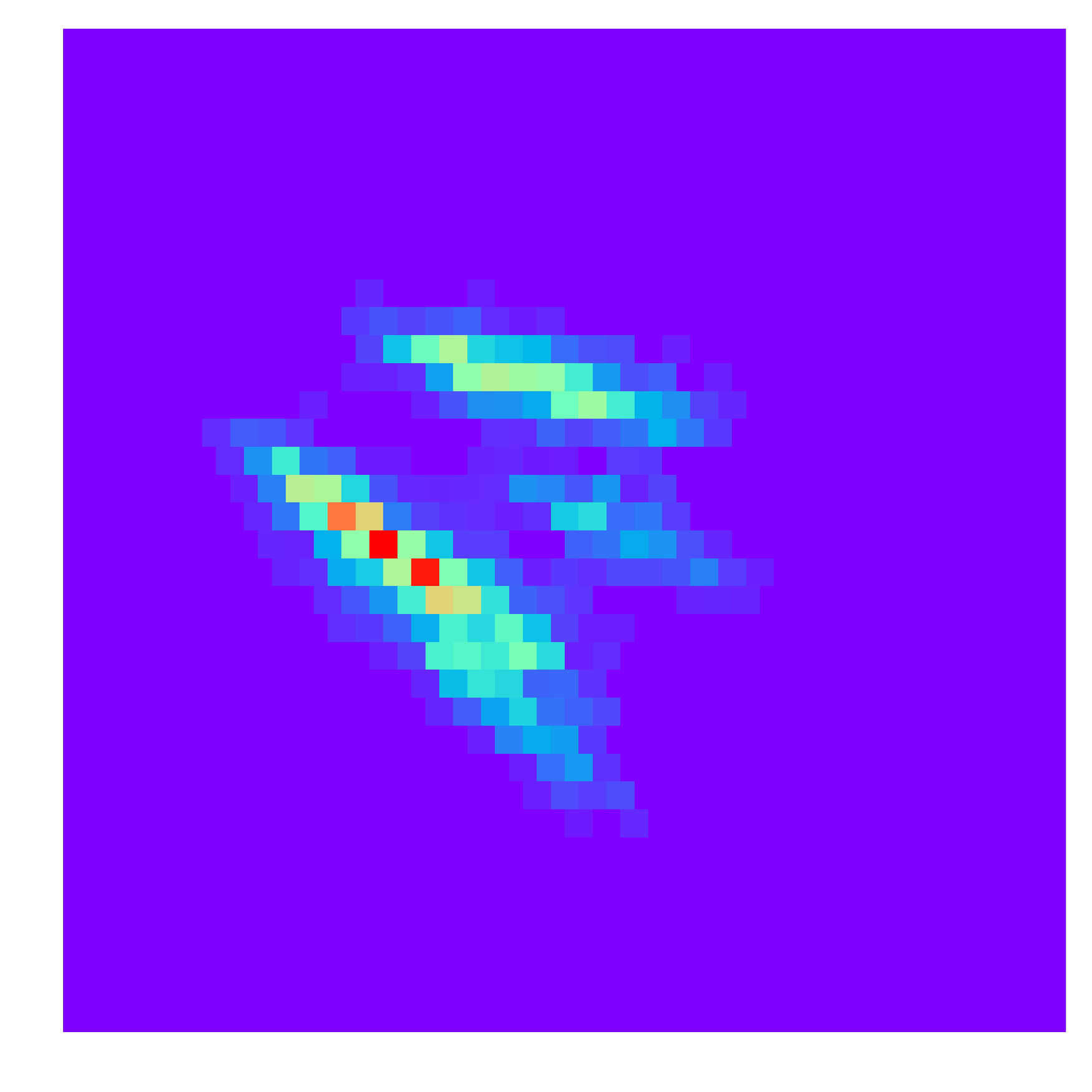}\\
\includegraphics[width=0.22\textwidth]{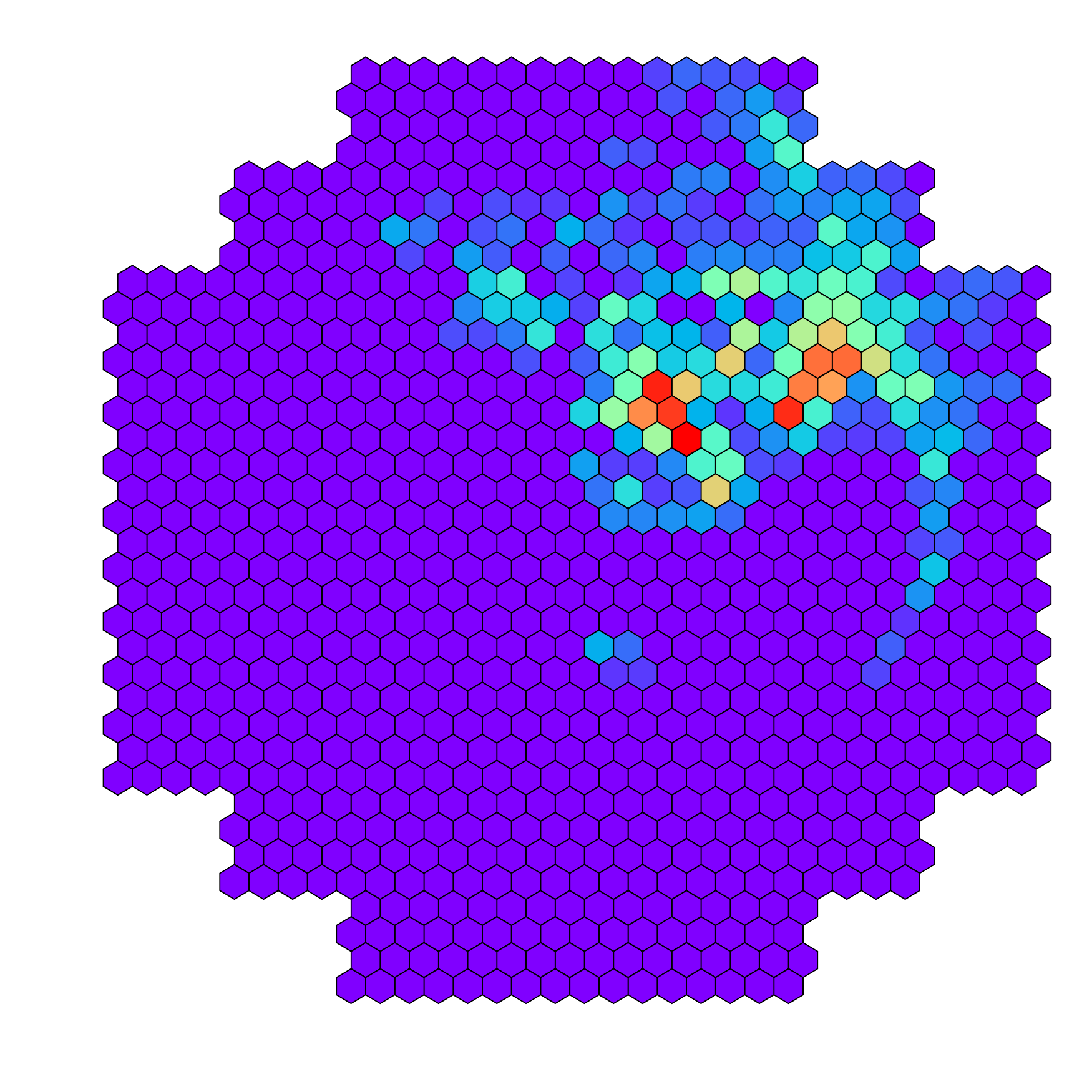}
\includegraphics[width=0.22\textwidth]{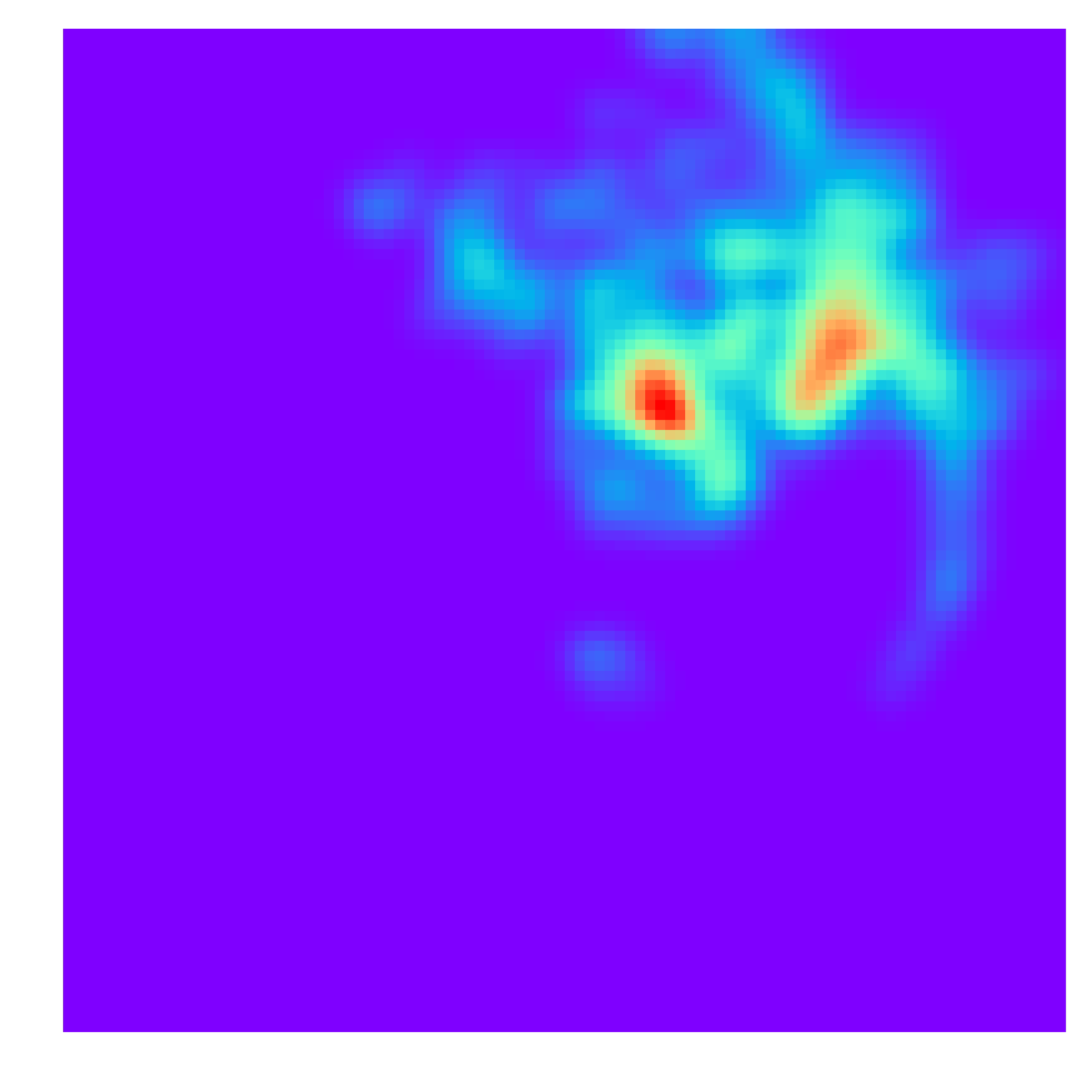}
\includegraphics[width=0.22\textwidth]{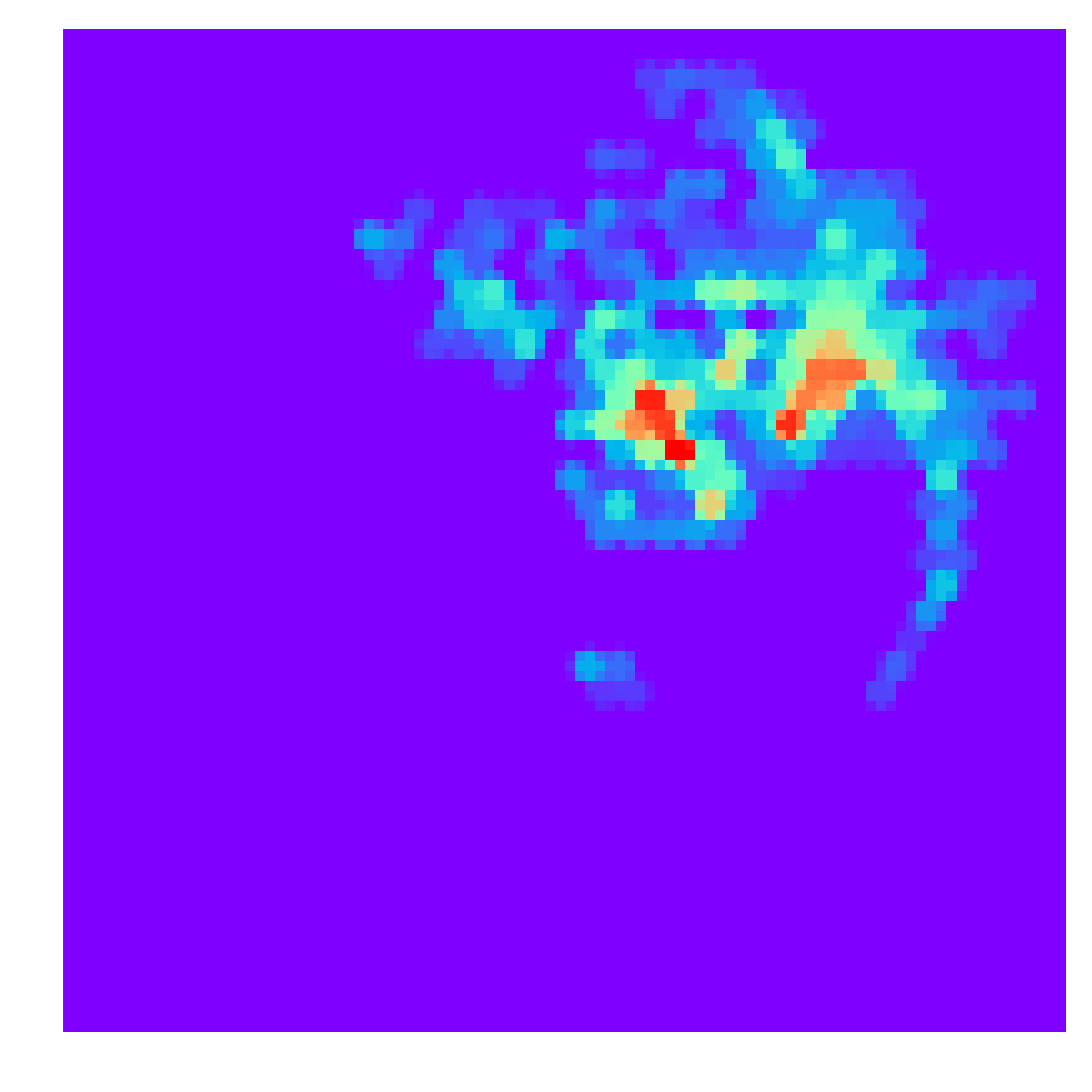}
\includegraphics[width=0.22\textwidth]{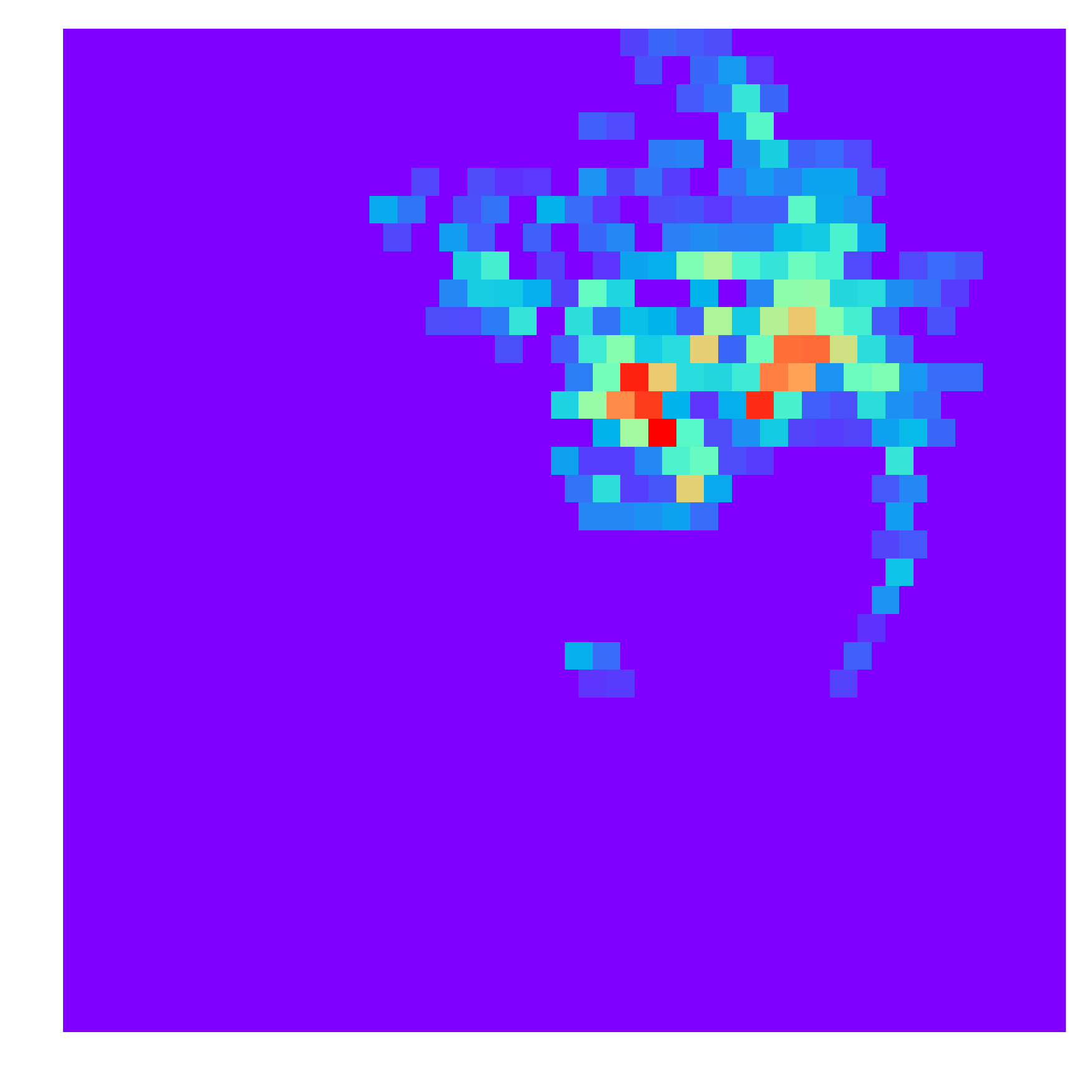}\\
\caption{Combined telescope images of a $\gamma$-ray (\textit{top row}) and a hadronic (\textit{bottom row}) event. \textit{Left column}: Camera pixel intensities. The square images show the same event sampled with Gaussian smoothing (\textit{middel left}), by rebinning (\textit{middle right}) and by oversampling (\textit{right}).}
\label{sampling_examples}
\end{figure}

To optimize the learning process of a CNN, one usually pre-processes the input data. Throughout this work we utilise images of simulated EAS events and select only events that trigger at least two of the CT1-4 telescopes. 
To incorporate stereoscopic information within a single image, in the first image processing step the $\mbox{CT1-4}$ images are combined into a single input image by projecting the individual images onto a common plane (see Fig.~\ref{sampling_examples}).

The H.E.S.S. cameras' pixels are arranged in a hexagonal grid. However, current DL frameworks, such as the ones used here, are designed to process inputs with a Cartesian grid. Therefore, we have developed a pre-processing chain to sample the EAS images to a square grid. This issue is not unique to IACTs, but a general problem of converting hexagonal sampled data to a square grid. A conversion from a hexagonal to a square lattice will always result in a change of topology, due to the different set of discrete symmetry groups that define the lattice. 

\begin{figure}
	\begin{center}
		\subfloat[high resolution rebin]{\includegraphics[width=0.2\textwidth]{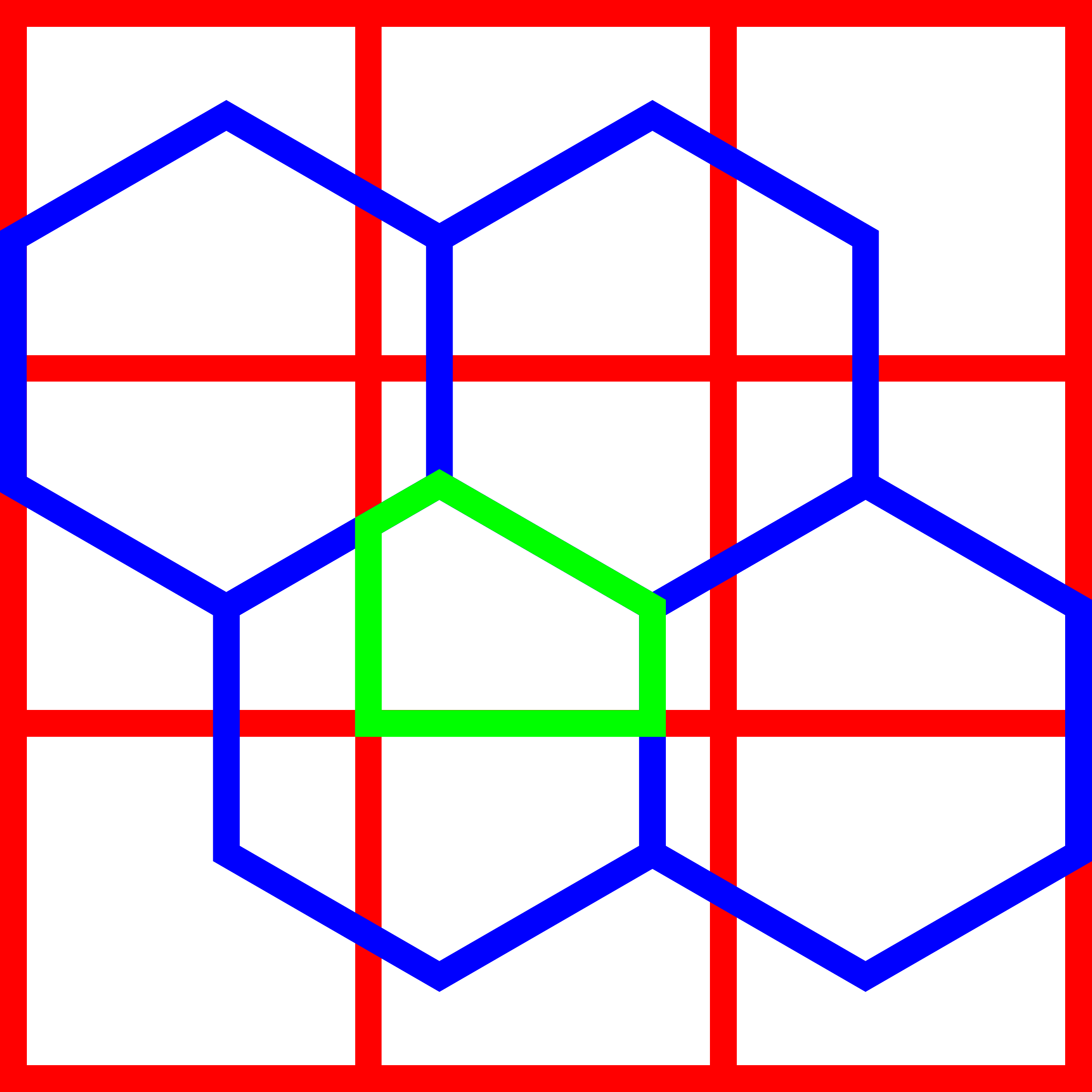}}
		\hspace{0.1\textwidth}
		\subfloat[low resolution rebin]{\includegraphics[width=0.2\textwidth]{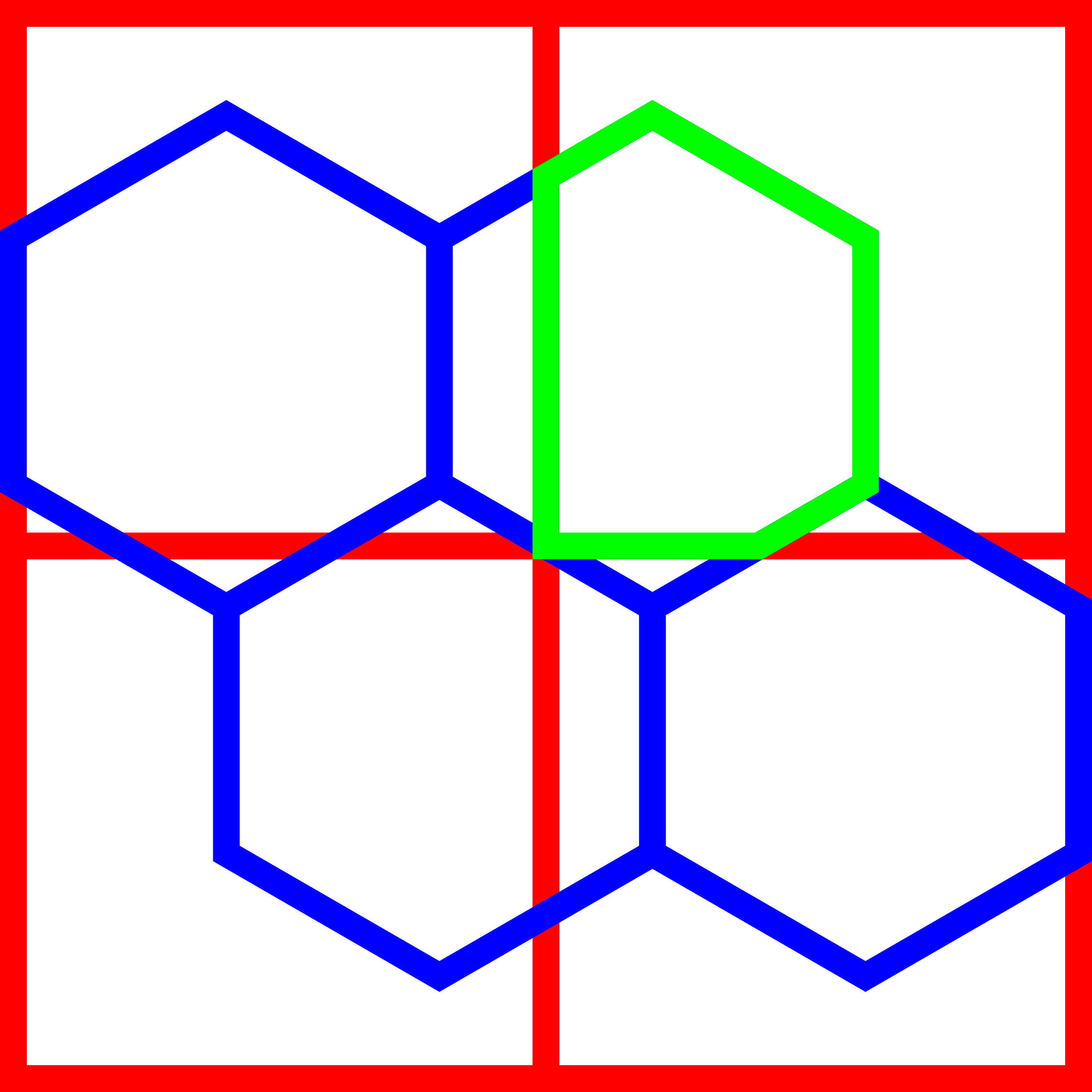}}
		\hspace{0.1\textwidth}
		\subfloat[oversampling]{\includegraphics[width=0.22\textwidth]{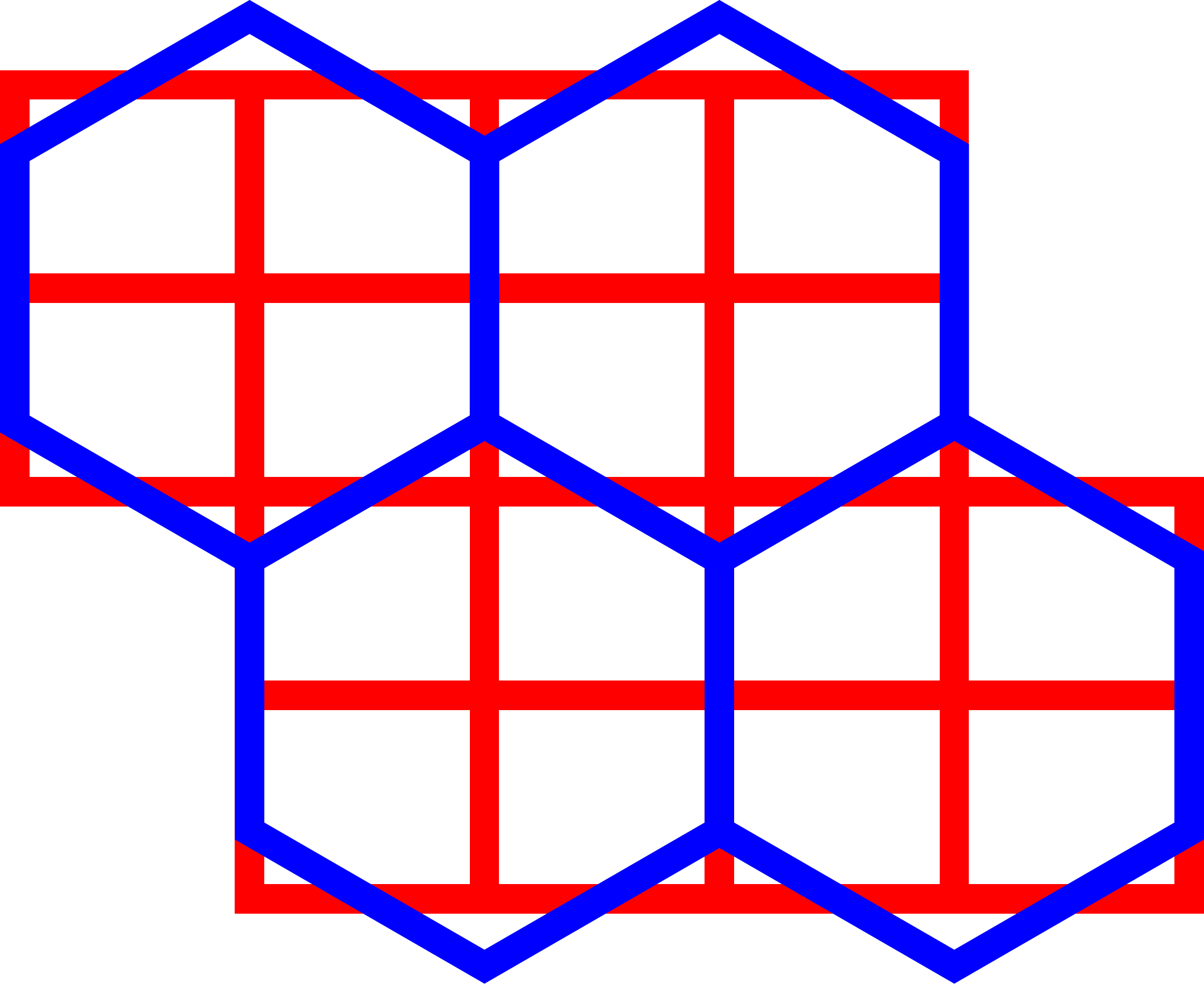}}
	\end{center}
	\caption{Visualization of two pre-processing methods. Panels \textit{(a)} and \textit{(b)} show our rebinning method, resampled with higher and lower resolutions (red squares), respectively. The \textit{green} line shows a portion of a hexagonal pixel that is attributed to a certain square pixel. Panel \textit{(c)} demonstrates the distortion effect caused by the oversampling method, where the red rectangles must be stretched in order to receive a square grid.} \label{fig:rebin_demo}
\end{figure}

We compared the resampling performance of several different approaches: standard linear and cubic interpolation methods, a straightforward oversampling method, a Gaussian smoothed sampling and a rebinning approach. The oversampling method, introduced by Feng and Lin in \cite{veritas_rebinning}, decomposes each hexagonal pixel into four rectangles, implying that the image is stretched in one direction. Gaussian smoothing first creates a histogram of pixel centers, weighted by intensity, and applies a Gaussian kernel, thereby smearing the contents. Interpreting camera pixels as histogram bins allows to rebin the image to a square histogram. An illustration of rebinning and oversampling is shown in Fig.~\ref{fig:rebin_demo}.

The comparison was done using a set of 10k random 2D Gaussian functions that resemble an elliptical EAS camera image. These analytical test functions were then sampled with a hexagonal grid, similar to the CT1-4 cameras' grid. Applying the pre-processing methods mentioned above, results in images of the desired square shape ($100\times100$ pixels for Gauss, rebinning and interpolation, $72\times72$ for oversampling). Fitting a Gaussian function to the square pixel images allows to compare the resampling methods in an analytical way. To describe the functions, we used a set of five shape parameters: center point coordinates (denoted $x$ and $y$), length ($\sigma_x$), width ($\sigma_y$) and orientation angle ($\theta$). The distributions of differences between fit- and function-parameters can be used to estimate the resampling methods' performance and are summarized in Table~\ref{samp}.


\begin{table}
\small
\centering
\begin{tabular}{ | c | d{2.2} @{${}\pm{}$} d{5.2}| d{2.2} @{${}\pm{}$} d{5.2} |d{3.2} @{${}\pm{}$} d{5.2} |}
\hline
  \multicolumn{1}{|c|}{Sampling Method}
& \multicolumn{2}{c|}{$|\Delta x|  \times 10 ^{-3}$} 
& \multicolumn{2}{c|}{$|\Delta \sigma_x| \times 10 ^{-3}$}
& \multicolumn{2}{c|}{$|\Delta\theta|  \times 10 ^{-3}$} \\
\hline\hline
Linear interpolation 	& 0.10 & 11.77 		& 0.14 & 16.78 		& 0.11 & 9.31  \\ 
Cubic interpolation 	& 0.05 & 11.78 		& 0.02 & 3.18 		& 0.04 & 4.66  \\
Oversampling 			& 42.04 & 10568.57 	& 11.49 & 5481.90 	& 101.40 & 17382.15 \\
Gaussian sampling 		& 4.02 & 390.99 	& 12.55 & 4326.78 	& 2.40 & 1174.06 \\
Rebinning 				& 0.07 & 27.73 		& 0.13 & 16.58 		& 0.09 & 9.52 \\
\hline
\end{tabular}
\caption{The distribution of the differences between the fit- and function-parameters for the resampling methods. $\Delta y$ and $\Delta \sigma_y$ show similar behavior to the $x$ coordinate and are therefore omitted.} 
\label{samp}
\end{table}

Table~\ref{samp} shows that cubic interpolation excels at shape conservation. However, it is by far the most computationally expensive method. In contrast, the Gaussian sampling and oversampling methods add large amounts of distortion to the images. Lastly, the performance of the rebinning and linear methods is comparable to that of the cubic interpolation. Nevertheless, because rebinning conserves the total image intensity, we generally prefer it over the linear interpolation.

Comparison of different resampling methods via artificial images should be taken merely as an estimate to rule out specific approaches. It is not a measure for actual network performance. Due to the statistical nature of machine learning based analyses, a resampling method should be chosen specifically to a given task and network architecture. Fig.~\ref{sampling_examples} shows examples of pre-processed data, resampled with our implemented Gauss, rebinning and oversampling methods.


\section{Data Sets and Training Results}

Our data sets were obtained from a set of MC data produced by simulating the interaction of both $\gamma$-rays and protons in the atmosphere with the CORSIKA software \cite{cors1} as well as the response of the H.E.S.S. telescopes with the sim\_telarray package \cite{corsika}. For the training, we have used $\gamma$-ray and proton EAS images from the latest H.E.S.S. simulations, denoted phase2b5. Both particle types were simulated as diffuse emission from 20$^{\circ}$ zenith and 180$^{\circ}$ azimuth, with a view cone of 5$^{\circ}$ and spectral index of $-2$. The energy distribution is 5$\,$GeV to 150$\,$TeV for $\gamma$-rays and  5$\,$GeV to 280$\,$TeV for protons. The images are cleaned according to the standard H.E.S.S. cleaning scheme \cite{image_cleaning}. Our {\it simulation set} consists of events that survived the image cleaning in at least two of the CT1-4 telescopes. CT5 images are omitted from this data set.

\subsection{Classification Tasks - Background Suppression}
\label{sec_classification}

To separate background from signal events, we trained a CNN to attribute an IACT event to one of two classes: $\gamma$-ray or proton. To create the data set for this task, we arbitrarily chose 640k events from our {\it simulation set}. The data set was then randomly divided into three sub-sets: a training set (60\% of the data set events), a validation set (20\%) and a test set (20\%). The ratio of $\gamma$-ray events to proton events in each of the sub-sets is 1 and the $\gamma$-ray energy range is 250$\,$GeV to 100$\,$TeV, while for protons it is 250$\,$GeV to 280$\,$TeV. To generate input images we use the rebin method with a resolution of 100$\,\times\,$100 pixels (due to the cameras' aspect ratio the true resampling resolution is 100$\,\times\,$97. To get a square image, we pad the image appropriately). 

An important point to notice is that no selection cuts are applied in order to select events for the sub-sets. This means, particularly due to the large energy range of the data set, that many of the events ($\sim$ 30\%) that the classifier was trained and tested on are truncated and would not have passed the pre-selection cuts of the H.E.S.S. analysis chain. In addition, no binning was applied to any parameter. Taking the binned training approach could, in principle, increase the accuracy score of the classifier. However, this means that when coming to analyse real data each event would have to be sent to its corresponding classifier. This requires knowledge about the particle properties (e.g. the energy for energy binned training) prior to the reconstruction stage. As the energy and direction of the $\gamma$-ray are not necessarily known at the classification stage, this approach was not favoured. 

The architecture of our classification network contains six learned layers - three CL and three FC layers. This architecture, together with the image pre-processing mentioned above, achieved the highest accuracy score out of several other architectures which combined 2-4 CLs with 2-3 FC layers and 100 $\times$ 100 pixels images. To control overfitting, we regularize each FC layer by applying weight decay with a constant of 0.004. The model was trained on a machine with two Nvidia\textsuperscript{TM} GeForce GTX 1080 GPUs. We took the data parallelization approach to accelerate the training process. 

\begin{figure}
\centering
\includegraphics[width=0.405\textwidth]{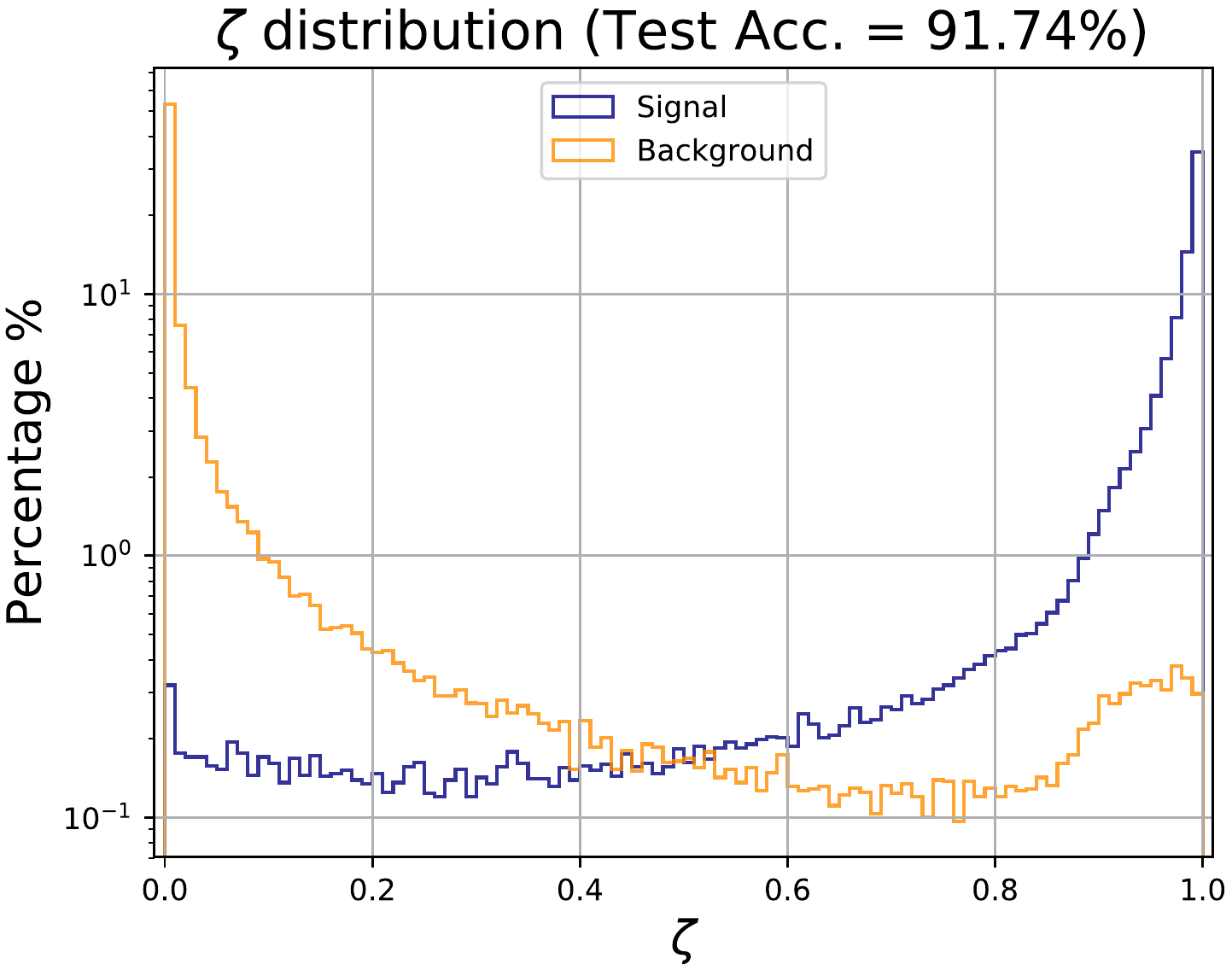}~~~~~~~~~~\includegraphics[width=0.4\textwidth]{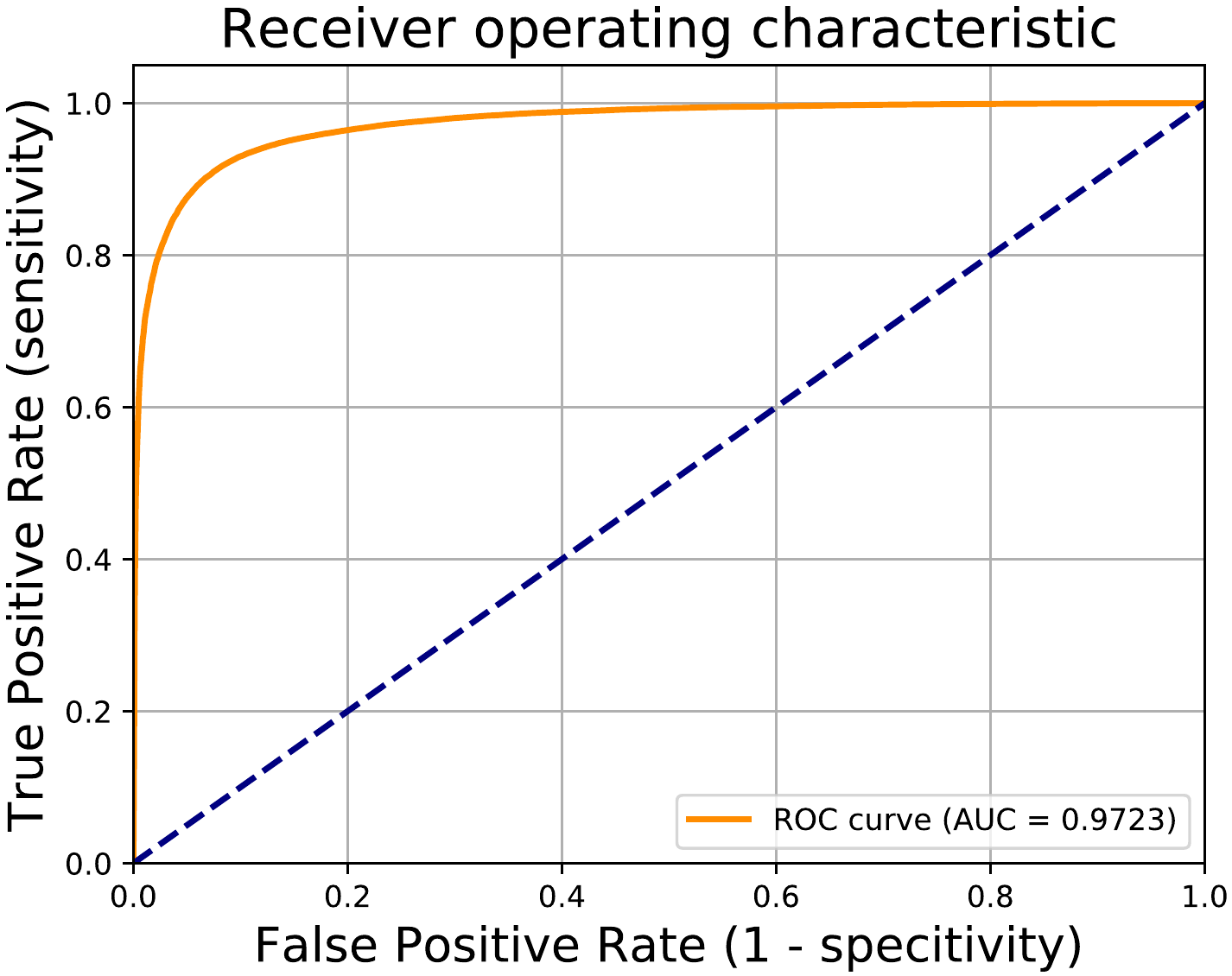}
\caption{$\zeta$-score distribution (\textit{left}) and ROC curve (\textit{right}) for the background separation task.}
\label{zeta}
\end{figure}

The test set is used to demonstrate how well the classifier generalizes to an arbitrary set of events from the simulation data (excluding events that were used for training and validation). Hence, the classification results presented here are obtained on the test set. To calculate our test accuracy, we used a $\zeta$ threshold of 0.5, where $\zeta$ denotes the output of the softmax function for a single event. An event is classified as a $\gamma$-ray when it receives a $\zeta$-score > 0.5 and otherwise as a proton ($\zeta$ can be understood as the "probability to be a $\gamma$-ray"). For this threshold, our network achieves a test set total accuracy of {\bf 91.7\%} (where total accuracy accounts for both $\gamma$-rays and protons which are correctly classified). We illustrate the general performance of the classifier by the ROC curve. To compare classifier performance, we use the area under the curve (AUC) metric. The $\zeta$ distribution and the ROC curve, for the complete test set, are shown in Fig.~\ref{zeta}. 

To learn the classifier's response to specific events, we calculated the accuracy for events in different energy and average local distance ranges. The local distance is given for each telescope image and is defined as the distance between the barycenter of the image to the camera center. The average local distance is calculated for the CT1-4 images of each event. The current H.E.S.S. analysis scheme omits images with local distance~>~2$^{\circ}$. The results are shown in Table~\ref{t1}.

\begin{table}
            \centering
            \begin{tabular}{| c | c ? c | c | }\hline
                Energy  & Test acc. & Avg. loc. dis.  & Test acc. \\\hline\hline
                0.25 - $1\,$TeV & 92.2\% & < 2$^{\circ}$ & 93.0\% \\ 
                1 - $10\,$TeV & 92.3\% & > 2$^{\circ}$ & 88.9\%  \\
                10 - $60\,$TeV & 89.6\% & - & - \\
                60 - $100/280\,$TeV & 90.7\% & - & - \\\hline
            \end{tabular}
            \caption{Test accuracies, evaluated in several energy ({\it left} columns) and avg. loc. dis. ranges ({\it right} columns). In the highest energy range, the energy maxima of the $\gamma$-rays and the protons are different and are $100\,$TeV and $280\,$TeV, respectively.
            The definition of an average local distance is given in the text. Images with local distance > 2$^{\circ}$ are usually truncated at the edge of the camera.} \label{t1}

    \end{table}

\subsection{Regression Tasks - Predicting Particle Properties}
\label{sec_regression}

\begin{figure}
	\centering
	\includegraphics[width=0.33\textwidth]{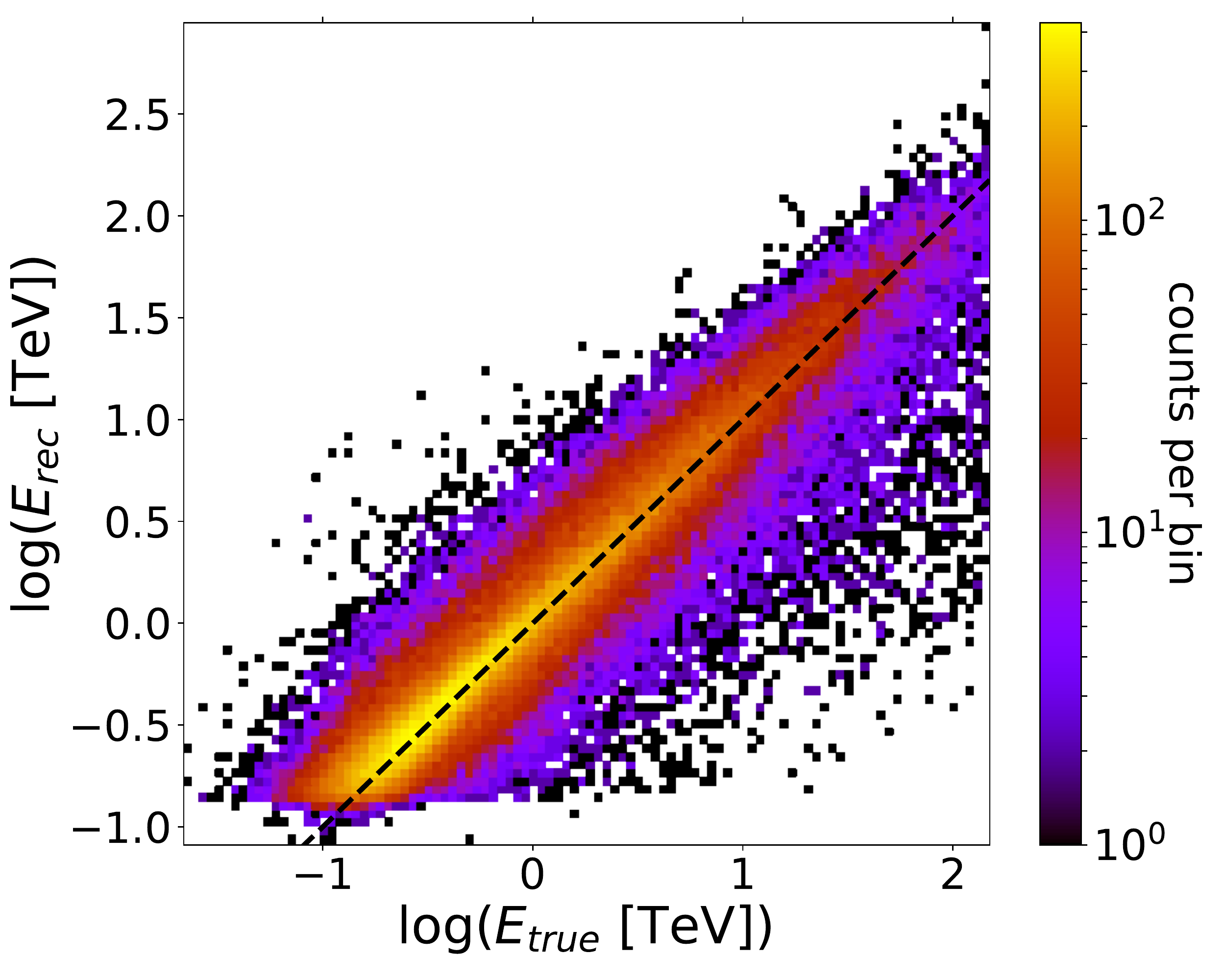}~~\includegraphics[width=0.33\textwidth]{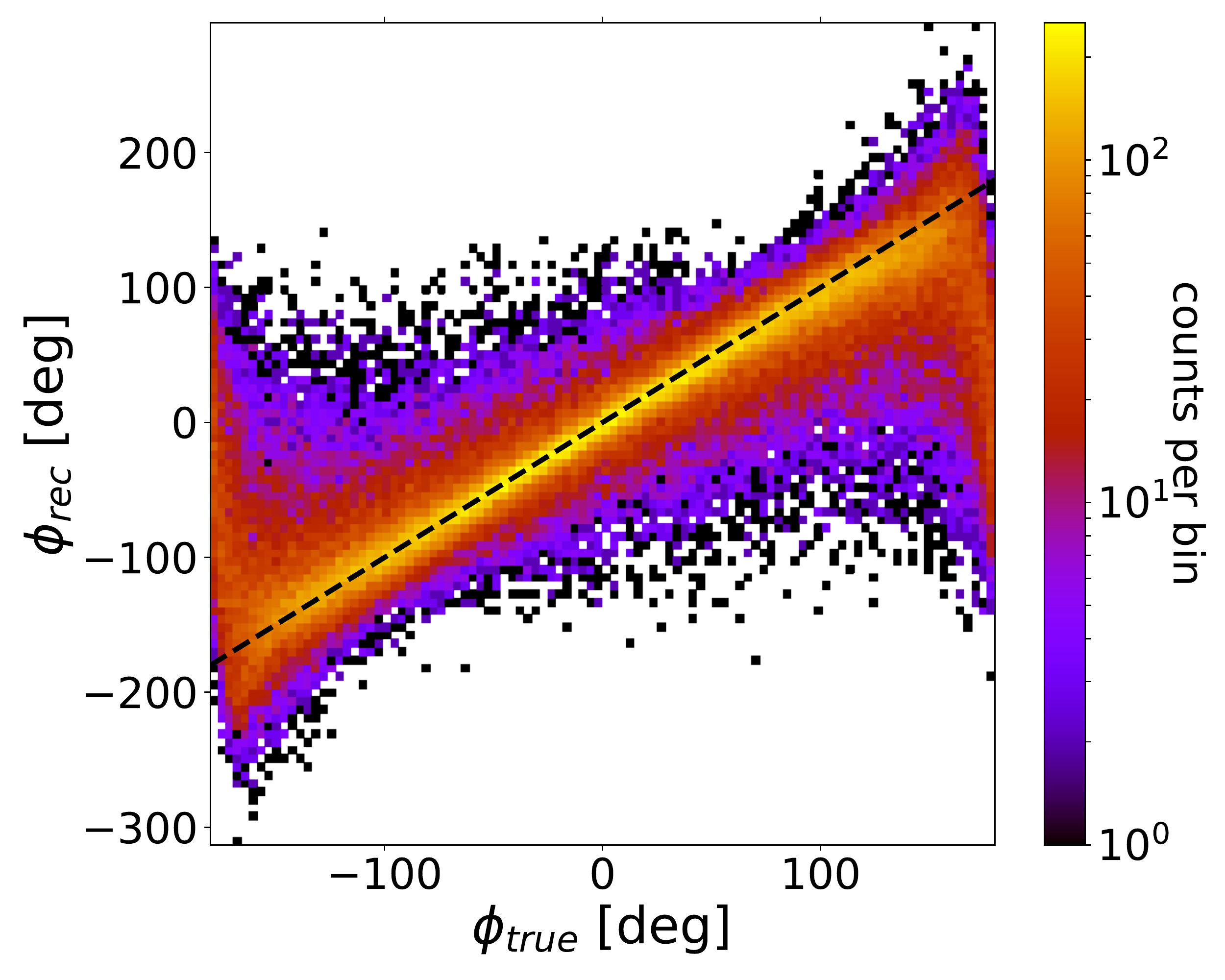}~~\includegraphics[width=0.33\textwidth]{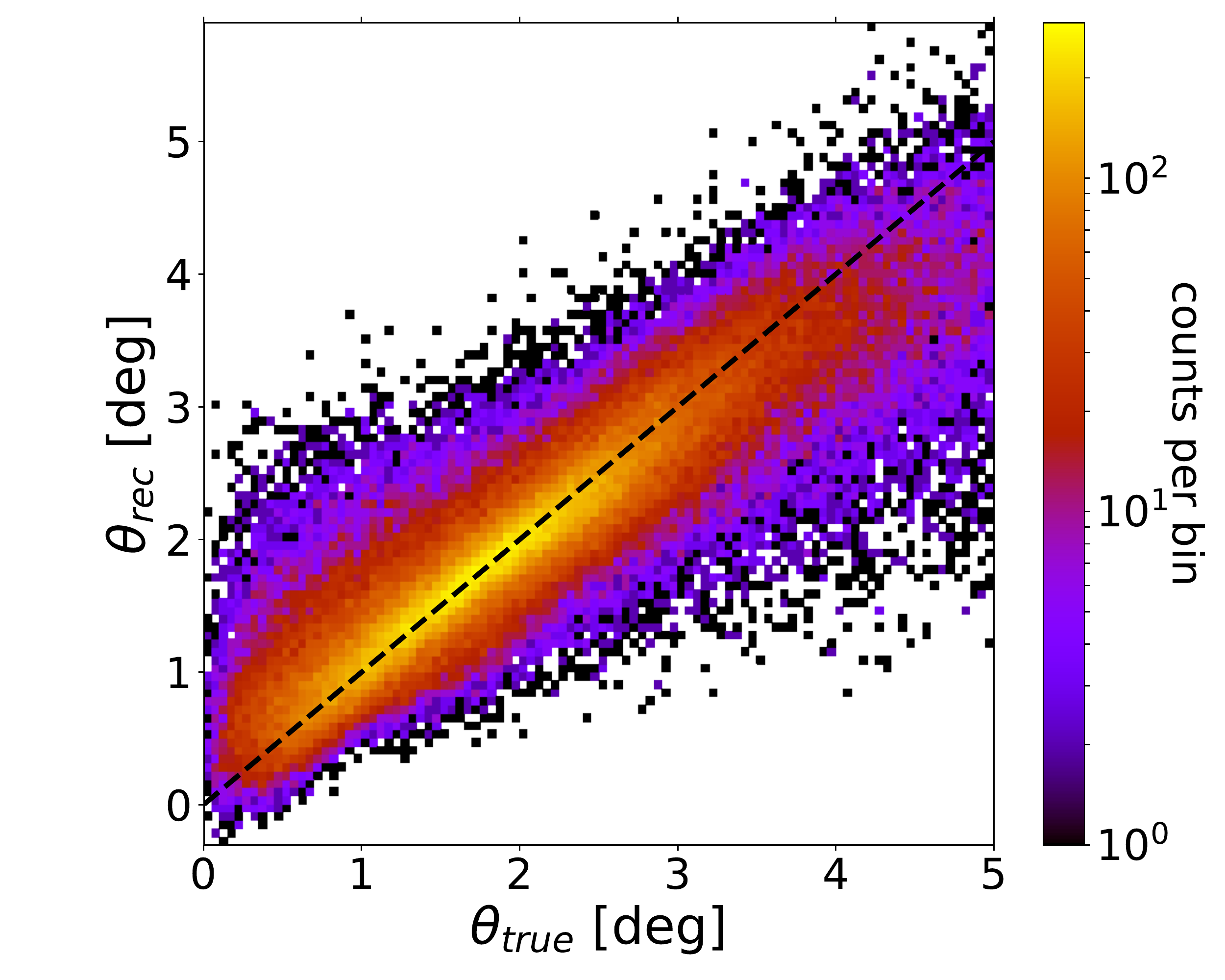}
	\caption{\textit{Left}: Migration matrix of true and reconstructed energy $E$. \textit{Center \& right}: Migration matrix of true and reconstructed event angle $\phi$ in the camera plane and offset $\theta$ from the camera center respectively. 
	}
	\label{fig_reg}
\end{figure}

A neural network trained for a regression task learns to output a continuous value instead of a discrete class value and thus can be used to predict properties of a primary shower particle. We have performed baseline studies on such regression problems by training similar CNN architectures as used for the classification with four CLs on $160$k random events from our simulation data set to predict $\gamma$-rays' direction and energy.
The energy of a $\gamma$-ray is strongly correlated to the total intensity of a shower image. A CNN is not able to propagate this information through its layers, but only extracts local characteristic intensity gradients which disassemble the total intensity information. Nevertheless, our trained CNNs are able extract enough information from the characteristic shapes in the images to model the $\gamma$-ray energy as seen in Fig.~\ref{fig_reg} (\textit{left}).
The direction of a $\gamma$-ray, on the other hand, is encoded in the translation of the intensity distributions in the camera plane. The intrinsic translational invariance of a CNN, however, does not prevent the network from modelling the direction information, as Fig.~\ref{fig_reg} (\textit{center \& right}) show.

These initial studies suggest that it is possible for CNNs to extract information from shower images to make predictions about the primary particle's properties on a level far from random guessing. CNNs may not be sensitive to the full set of prominent features of a property which are directly accessible to other reconstruction algorithms.

\section{Conclusion \& Outlook}

This work demonstrates that CNNs could be applicable in IACT data analysis. The pre-processing of the image data as well as the design of the network architectures can strongly influence the behaviour of the analysis and have to be thoroughly studied and tuned to reach optimal performance.
We have shown that CNN-classifiers can accurately separate IACT signal from background, yielding high accuracies of up to $91.7\%$ for a data set consisting of diffuse events with a continuous energy spectrum without applying any selection cuts. This implies that a significant number of events, that would have been rejected by commonly used selection cuts, can be further processed. 
Furthermore, we found that CNNs can be used to extract primary particle properties, i.e. energy and direction, from shower images but are probably limited by the actual patterns characterizing these properties. We will perform detailed studies on that matter in the future and evaluate the potential of applying CNNs in full shower reconstructions.

To improve our models' performance in regions of poor statistics (i.e. towards the upper and lower energy threshold, as well as towards low offsets from the camera center) we will create balanced data sets to counteract these low statistics at the boundaries. For future studies we also plan to combine data from different telescope types by applying so-called concatenation layers through which multiple CLs trained on different input data sets can produce a combined output.

The outstanding performance that CNNs exhibit in computer vision has promoted them to be used in many different commercial and scientific applications. Our work demonstrates a successful proof-of-concept for the application of DL techniques in $\gamma$-ray astronomy.

\acknowledgments

The authors would like to thank the H.E.S.S. Collaboration for the support on this research study and for granting access to the MC simulation data. The authors also thank Christopher van~Eldik and Alexander Ziegler for encouraging discussions.

\end{document}